\UseRawInputEncoding
\documentclass[
 reprint,
 amsmath,amssymb,
 aps,
 superscriptaddress
]{revtex4-2}

\usepackage{graphicx} 
\usepackage{dcolumn} 
\usepackage{bm} 
\usepackage{mathrsfs}
\usepackage{subfigure}

\begin{document}
\preprint{APS/123-QED}

\title{Neural Network Model for Structure Factor of Polymer Systems}
\author{Jie Huang}
\affiliation{Department of Physics, Wenzhou University, Wenzhou, Zhejiang 325035, China}

\author{Xinghua Zhang}%
\email{zhangxh@bjtu.edu.cn}
\affiliation{School of Science, Beijing Jiaotong University, Beijing 100044, China}

\author{Gang Huang}
\affiliation{Institute of Theoretical Physics, Chinese Academy of Sciences, Beijing 100190, China}

\author{Shiben Li}
\email{shibenli@wzu.edu.cn}
\affiliation{Department of Physics, Wenzhou University, Wenzhou, Zhejiang 325035, China}

\date{\today}

\begin{abstract} 
As an important physical quantity to understand the internal structure of polymer chains, the structure factor is being studied both in theory and experiment. Theoretically, the structure factor of Gaussian chains have been solved analytically, but for wormlike chains, numerical approaches are often used, such as Monte Carlo (MC) simulations, solving modified diffusion equation (MDE), etc. In those works, the structure factor needs to be calculated differently for different regions of the wave vector and chain rigidity, and some calculation processes are resource consuming. In this work, by training a deep neural network (NN), we obtained an efficient model to calculate the structure factor of polymer chains, without considering different regions of wavenumber and chain rigidity. Furthermore, based on the trained neural network model, we predicted the contour and Kuhn length of some polymer chains by using scattering experimental data, and we found our model can get pretty reasonable predictions. This work provides a method to obtain structure factor for polymer chains, which is as good as previous, and with a more computationally efficient. Also, it provides a potential way for the experimental researchers to measure the contour and Kuhn length of polymer chains. 
\end{abstract}

\maketitle

\section{\label{sec:level1}Introduction}
The structure factor of a polymer system defined as 
\begin{equation}
S(\mathbf{k})=\frac{1}{\rho} \int_{V}\langle\rho(\mathbf{r}) \rho(0)\rangle \exp (i\mathbf{k} \cdot \mathbf{r}) \mathrm{d} \mathbf{r}
\end{equation}
 is a measurable physical property, which characterizes the density–density correlation of the system.\cite{Svergun1987}  In theory, the structure factor can be used in field theory calculations. In the Gaussian fluctuation theory \cite{zhang_a2, Zhang2014}, the structure factor of the interacting system is calculated using the structure factor of the ideal chain. Besides, in the dynamic mean-field theory, the inter-chain correlation properties in the diffusion process \cite{zhang_a3, zhang_a4, zhang_a5} are also described by the structure factor of the ideal chain. Experimentally, the basic characteristics of polymers such as the degree of polymerization, the rigidity, and the chirality can be analyzed by fitting the scattering data with the structure factor.  As for calculation, the structure factor can be predicted from a microscopic chain model. The well-known expression of the Gaussian chain model has a Debye function form\cite{M.Doi;S.F.Edwards1986} and can be used to analyze the experimentally determined structure factor for a $\theta$-point dilute polymer solution in a moderate to small wavenumber range, $ka \leq 1$, where $a$ is the Kuhn length.

Besides, there is a large class of semiflexible polymer chains, where the effects of finite rigidity are important, which cannot be described by the Gaussian chain model. The wormlike chain model is one of the best semiflexible chain model. In this model, the polymer is an inextensible thread subject to a linear-elastic bending energy.\cite{Landau1986} The configuration of a wormlike chain of total length $L$ is described by a smooth space curve with its coordinate specified by $\mathbf{R}(s)$, where s is an arc-variable continuously varying from one end ($s = 0$) to another ($s = 1$).\cite{M.Doi;S.F.Edwards1986,Saito1967,KarlF.Freed1972} The Boltzmann weight for such a configuration is given by
\begin{equation} \label{eq:1}
\mathscr{W}[\mathbf{R}(s)]=\exp \left[-\beta H_{0}\right]
\end{equation}
where
\begin{equation} \label{eq:2}
\beta H_{0}=\frac{a}{4 L} \int_{0}^{1} \mathrm{d} s\left|\frac{\mathrm{d} \mathbf{u}(s)}{\mathrm{d} s}\right|^{2}+\frac{L}{a} \int_{0}^{1} \mathrm{d} s w[\mathbf{R}(s), \mathbf{u}(s)]
\end{equation}
The tangent vector $\mathbf{u}(s) \equiv(1 / L) \mathrm{d} \mathbf{R}(s) / \mathrm{d} s$ specifies the local orientation of the polymer chain at location $s$.  $u(s)$ is a unit vector, and $|u(s)| = 1$ due to the local inextensible constraint. The first term describes an energy penalty for a bent curve. Originally, a bending energy modulus $\beta \varepsilon$ was written as the coefficient; \cite{Saito1967} upon identification of the free-space mean-square radius of gyration with that of a Gaussian chain in the large $L/a$ limit, we can show that the prefactor can be written in the current form, where
\begin{equation}
a=2 \beta \varepsilon
\end{equation}
for a three-dimensional system. The Kuhn length $a$ is directly used here for comparison with results calculated from a Gaussian-chain model. The wormlike chain model involves two characteristic length scales: the length of chain $L$, and the effective Kuhn length $a$. 

The key to calculating the structure factor is the calculation of the Green's function (with $w=0$) in  Eq.\ref{eq:2}. As it turns out, no analytic expression of the Green's function is available for the wormlike chain model, as indicated earlier by Stepanow\cite{Stepanow2004, Stepanow2005}, Spakowitz and Wang\cite{Spakowitz2004} as well as by  Zhang\cite{Zhang2014, zhang_a1}.

Kholodenko exploited the similarity between the Green's function of the semiflexible polymer model and the propagator of a Dirac's fermion, in rigid and flexible limits.\cite{Kholodenko1993a, Kholodenko1990}  The limits for Gaussian-chain and rod expressions can be reproducible from the formula. It is by far the simplest, in comparison with the approximations proposed earlier by Yoshizaki and Yamakawa\cite{Yoshizaki1980} and later by Pedersen and Schurtenberger.\cite{Pedersen1996a} 

Pedersen and Schurtenberger performed a series of Monte-Carlo simulations of such a chain, with and without the excluded-volume interaction between monomers. The structure factor can then be obtained numerically from the simulations.\cite{Pedersen1996a} They have provided an empirical formula to represent their simulation data. More recently, Hsu and coworkers calculated the structure factor of a semiflexible chain model on a simple cubic lattice, using Monte Carlo simulations.\cite{Hsu2012, Hsu2013} 

Spakowitz and Wang proposed an alternative approach; calculating the problem of constrained one-dimensional random walk, they obtained the Green's function of a wormlike chain formally.\cite{Spakowitz2004} They re-grouped the random walk trajectories according to the number of loops in a loop expansion of the problem. Based on this consideration, the moment expansion can be expressed as an infinite continued fraction. The calculation of the continued fraction problem is equivalent to inverting a matrix that has the same format as the matrix used in Stepanow's work. To find the structure factor, however, one must go back to the numerical treatment of the formalism; in particular, an inverse numerical Laplace transformation is needed.\cite{Spakowitz2004, Mehraeen2008}

In our previous work\cite{Zhang2014}, a numerical method to obtain the structure factor of a homogenous wormlike polymer solution, based on the standard wormlike chain model was obtained. We calculated the $s$-dependent Green's function, utilizing a formal solution to the modified diffuse equation(MDE)\cite{KarlF.Freed1972,Liang2013} that the Green's function in Fourier space satisfies, and propagating the solution as $s$ increases. This method was numerically more straightforward than some other approaches suggested recently. And the solution captured the correct physical behavior of the structure factor in the entire parameter space of $L/a$ and $ka$. 

The motivation of this work is twofold. First, we attempt to find a more efficient formulation of structure factor for wormlike chains in the entire parameter space of $L/a$ and $ka$,  but the more direct way where we don't need to do any heavy calculation like Monte Carlo simulations or solve partial differential equations.  Second, we want to build a possible measure tool for the scattering experiments of polymer chains. If scattering intensity data is given, the contour length $L$ and Kuhn length $a$ can be easily obtained.

Recently, neural networks (NNs), as an important branch of machine learning(ML), are widely used in polymer physics, such as classifying phases of matter\cite{VanNieuwenburg2017}, solving nonlinear partial differential equations(PDE)\cite{Raissia}, predicting the structure of macromolecules\cite{Li2019a}, and polymer conformations classification\cite{Vandans2020a}.  Since NN has been proven to be able to approximate almost any functions\cite{Cybenko1989}, we do not need to find structure factor from the perspective of guessing the analytic formula, but only need to use a NN to replace it. To ensure the high accuracy of a NN model, a sufficient data set is needed to train this NN. Fortunately, we can get numerous exact structure factor data with different $ka$ and $L/a$ by using the method in Ref.\cite{Zhang2014}.  By training with a data set, we can get a trained NN model to calculate the structure factor easily. 

The outline of the rest of the paper is as follows. We first demonstrate how to apply NN to structure factor fitting in section \ref{sec:level1}, including the basic introduction of NNs, the training task, the training process, and the influence of NN architecture.  Following this, we build a method to predict the contour length $L$ and Kuhn length $a$ of polymer chains in section \ref{Sec.2.P} by using the trained NN.

\section{\label{sec:level1}The neural network model for structure factor} 

\subsection{ \label{subsec:IntroNN}A brief introduction to  Neural Networks }

 \begin{figure}[b]
	\centering 
	\subfigure[]
	{
		\includegraphics{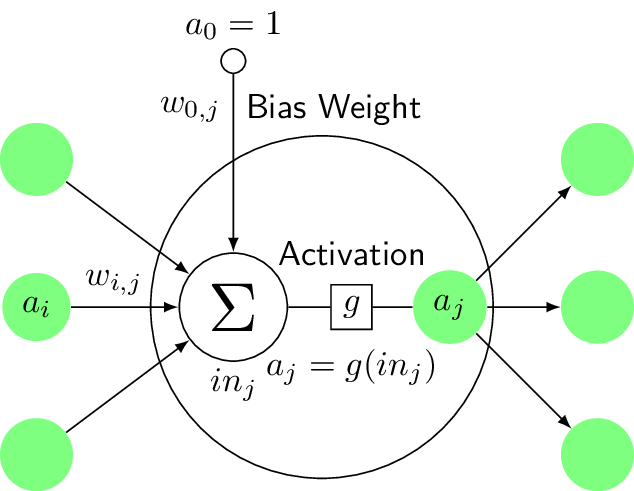}%
		\label{fig:neuron}%
	}
	
	\subfigure[]
	{
		\includegraphics{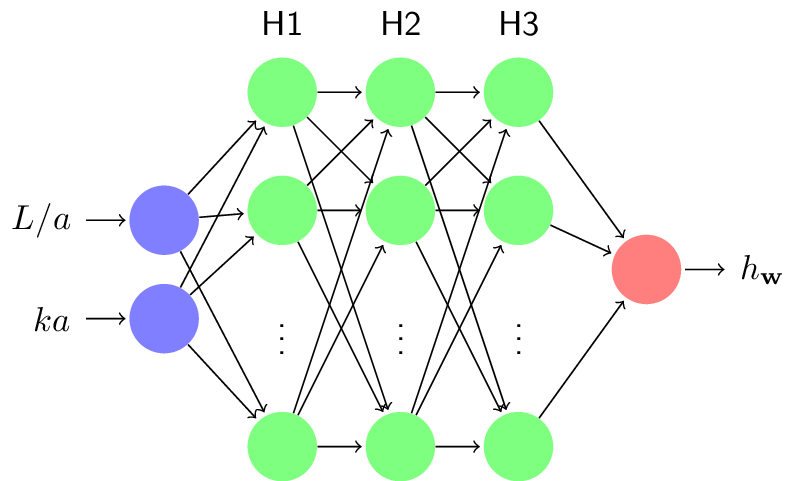}%
		\label{fig:nn_structure}%
	}
	\caption{(a) The mathematical model for a neuron. The unit's output activation is $a_j = g(\sum_{i=0}^nw_{i,j}a_i)$, where $x_i$ is the output of unit $i$ and $w_{i,j}$ is the weight on the link from unit $i$ to this unit; (b) A fully connected NN with 3 hidden layers.} 
	\label{fig:neuron_nn}%
\end{figure}

To begin with, we introduce some basic concepts of neural networks. NNs, also called artificial neural networks(ANNs), are computing systems which can learn to perform different tasks by considering example generally without being programmed with task-specific rules.  A NN is based on a collection of connected nodes or units called \emph{neurons}. As in Fig.~\ref{fig:neuron},   a link from neuron $i$ to neuron $j$ serves to propagate the activation $a_i$ from $i$ to $j$. Each link also has a numeric weight of $w_{i,j}$ associated with it, which determines the strength and sign of the connection. And each neuron has a dummy input $a_0=1$ with an associated weight $w_{0,j}$. Each neuron $j$ first computes a weighted sum of its inputs:

\begin{equation}
in_j = \sum_{i=0}^nw_{i,j}a_i.
\nonumber
\end{equation} 
Then it applies an activation function $g$ to this sum to derive the output\cite{Norvig2010a}:
\begin{equation} \label{nn_cal}
a_j = g(in_j) = g\left( \sum_{i=0}^nw_{i,j}a_i \right).
\end{equation}
The nonlinear activation function is the key to the power of NN that allows it to approximate almost any function.  In this work, the sigmoid function was used:
\begin{equation}
g(z)=\frac{1}{1+e^{-z}}.
\nonumber
\end{equation}

Having decided on the mathematical model for the individual neurons, then a fully connected NN can be obtained by connecting them. As shown in Fig.~\ref{fig:nn_structure},  a fully connected NN is arranged in layers, which can be divided into an input layer, many hidden layers, and an output layer. Only the input layer doesn't participate in the calculation in Eq. \ref{nn_cal}. There can be many neurons on each layer. Any neuron in the hidden layer and output layer is connected to all neurons in the previous layer.  

A NN can be viewed as a mapping $h$ from its input $\mathbf{x}$ to its output $h_{\mathbf{w}}(\mathbf{x})$, where $\mathbf{w}$ is the collection of all the weights in this NN. By changing $\mathbf{w}$, different $h$'s can be obtained.  The process of tuning $\mathbf{w}$ to approximate another function $f$ is called \emph{training} the NN.  And we train the NN by showing lots of input-output pairs repeatedly to the NN so that it can gradually \emph{learn}  the mapping from the input to output by tuning the  $\mathbf{w}$. The input-out pairs constitute a \emph{training set}, and this type of learning is called \emph{supervised}  learning. 

To be more specific, the training task can be described as follows. Given a training set of N example input-output pairs $(\mathbf{x}_1,y_1), ... , (\mathbf{x}_j,y_j) , ... , (\mathbf{x}_N,y_N)$,  where $\mathbf{x}_j = ((L/a)_j, (ka)_j)^T$,  and $y_j$ was generated by the method in \cite{Zhang2014}
\begin{equation}
y = f(\mathbf{x}) =  (L/a){(ka)}^2S(L/a, ka),
\end{equation}
find a function $h$ that approximates the function $f$. The way to train the NN is by following.  At first, we defined a loss function, 
\begin{equation}
Loss( \mathbf{w} )  = \frac{1}{N}\sum_{x} \|f(x) - h_{\mathbf{w}}(x)\|^2
\end{equation}
which indicates how far away the  $h$ is from the objective function $f$. By training the NN, we'd like to  find the weights  $ \mathbf{w} ^* $ so that the loss function over the examples could be minimized, i.e., 
\begin{equation}
\mathbf{w^*} = \arg \min_{\mathbf{w}} Loss( \mathbf{w} ) .
\end{equation}

There are many optimization algorithms, also called \emph{optimizer}s, to find $\mathbf{w ^ *} $, but the basic idea  can be expressed as follows. The weight $w_i$ is updated by
\begin{equation}
w_i \leftarrow w_i - \alpha\frac{\partial{Loss(\mathbf{w})}}{\partial{w_i}}, 
\end{equation}
where $\alpha$ is the learning rate. In this work, we use an adaptive learning rate optimizer called Adam\cite{Kingma2015}   that  is  designed specifically for training NNs. To increase training efficiency, the training set is usually divided into many mini-batches. Each time a mini-batch is used to update $\mathbf{w}$.  Besides, the training set is used to update $\mathbf{w}$ many times. The process of updating $\mathbf{w}$ using a training set once is also called an \emph{episode}.

The training set for the structure factor of this work comes from the numerical method in \cite{Zhang2014} which obtained an excellent agreement between the structure factor computed from the method of infinite continuous fractions by Spakowitz and Wang\cite{Spakowitz2004}. 
Compared with other methods like the Dirac propagator approach\cite{Kholodenko1993a} or Monte Carlo simulations \cite{Pedersen1996a},  this method gives rise to the precise determination  of structure factor in the entire $ L/a$--$ka$ space ($ L / a, ka \in [10 ^ {-2}, 10 ^ {3}] $), especially in low and large $L/a$ regime, so that it laid the foundation for reliable training set.

In \cite{Zhang2014},  the polynomial $ (L / a) (ka) ^ {2} S $ is a function with $ L / a $ and $ ka $ as arguments. We consider the $\mathbf{x} = (L / a ,ka)^T $  and the corresponding $ (L / a) (ka) ^ {2} S $  as a training sample.  For each $L/a$, 100 points for $ka$ in  $[10^{-2},10^3]$ are uniformly sampled on $\text{ln} (ka)$.  Similarly, 100 points  for $L/a $ in $[10^{-2},10^3]$ are uniformly sampled on $\text{ln}(L/a)$. Therefore, 10,000 training samples were obtained, which covers the entire domain of  $ka$ and $L/a$ described by wormlike chain model.

\subsection{\label{sec:level6} Training results of the NN model}

By using Tensorflow, we chose Adam optimizer with a fixed learning rate of $10^{-5}$,  $4$ hidden layers,  and $25$ neurons per layer.  After  $ 6\times 10^5 $ epochs of training, the loss value $L$ was reduced to $6.9\times10^{-7}$. 

\begin{figure}
	\includegraphics{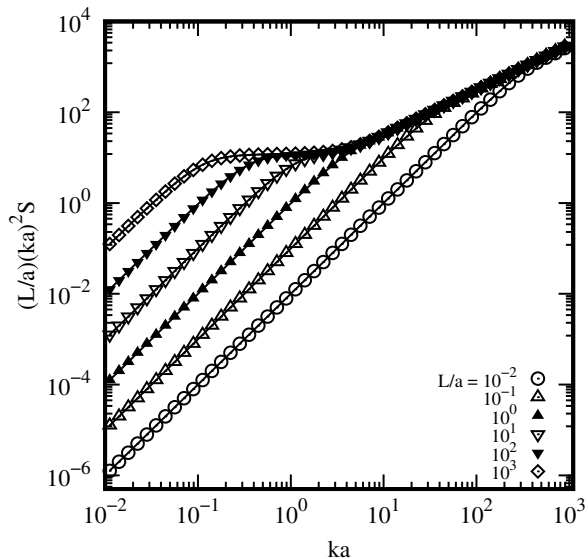}%
	\caption{\label{fig:fitting_compare_in_2D}Structure factor comparison between the target values (circles, triangles, etc) and  trained NN predictions (lines) in logarithmic coordinates with $4$ hidden layers and $25$ nodes on each hidden layer with $Loss = 6.92\times10^{-7}$.
	}
\end{figure}

To demonstrate the effectiveness of the structure factor in the form of NN, 
in Fig.~\ref{fig:fitting_compare_in_2D},  we've  plotted the NN model predictions of 5 different rigidities with  $L/a = 10^{-1}, 10^{0}, 10^1, 10^ 2$ and $10^3$.  For comparison, we also made the solution of the structure factor given by \cite{Zhang2014} denoted by circles, triangles, etc.  The NN can well represent the structure factor for different rigidities for the  entire $k$ range, and consistent with the exact result obtained by the method proposed by \cite{Zhang2014}.

\begin{figure*}
	\includegraphics{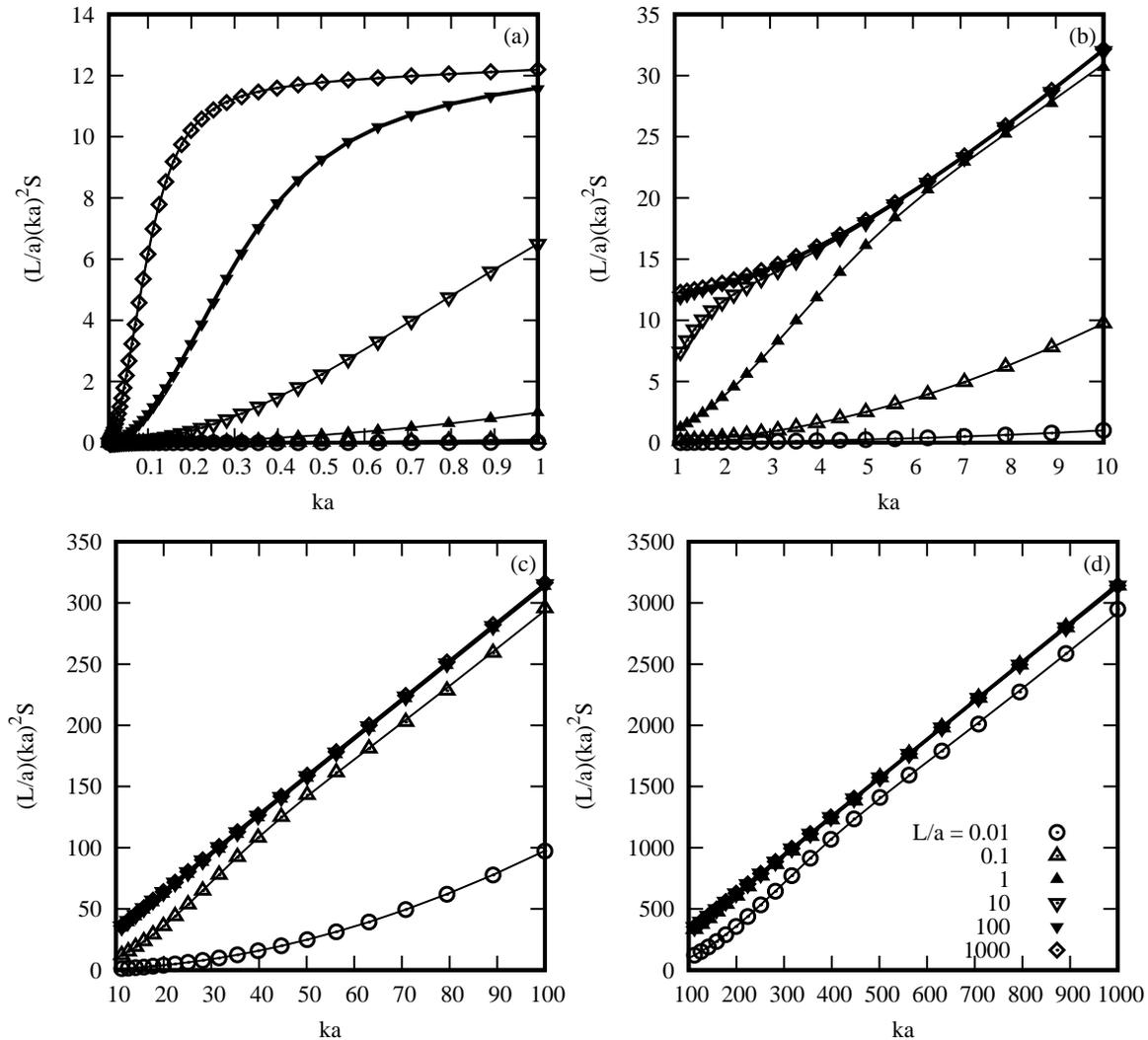}%
	\caption{\label{fig:fitting_compare_in_linear}Structure factor comparison  between the target values (circles, triangles,etc.) and  trained  NN predictions (lines) in linear coordinates. (a) $ka < 1$; (b) $ka \in [1,10]$; (c) $ka \in [10,100]$; (d) $ka \in [100,1000]$.
	}
\end{figure*}

To further verify the fitting results of the trained NN, we made comparisons at different scales in linear coordinates. As shown in Fig.~\ref{fig:fitting_compare_in_linear}, $a$, $b$,  $c$,  and $d$ respectively correspond to different $ ka $ regions $ [10 ^ {-2}, 10 ^ {-0}], [10 ^ 0, 10 ^ 1], [10 ^ 1, 10 ^ 2], [10 ^ 2,10 ^ 3] $, where the solid lines are the values given by the NN model, and the circles, triangles, etc are the solutions from \cite{Zhang2014}.  It can be concluded  that  the NN model can give highly accurate structure factor values in the \emph{entire} $L/a-ka$ space.  Therefore, our model also has a good description of rigid and semi-rigid polymer chains, which is of practical significance, such as fluctuation theory and scattering experiment.

\begin{figure}
	\includegraphics{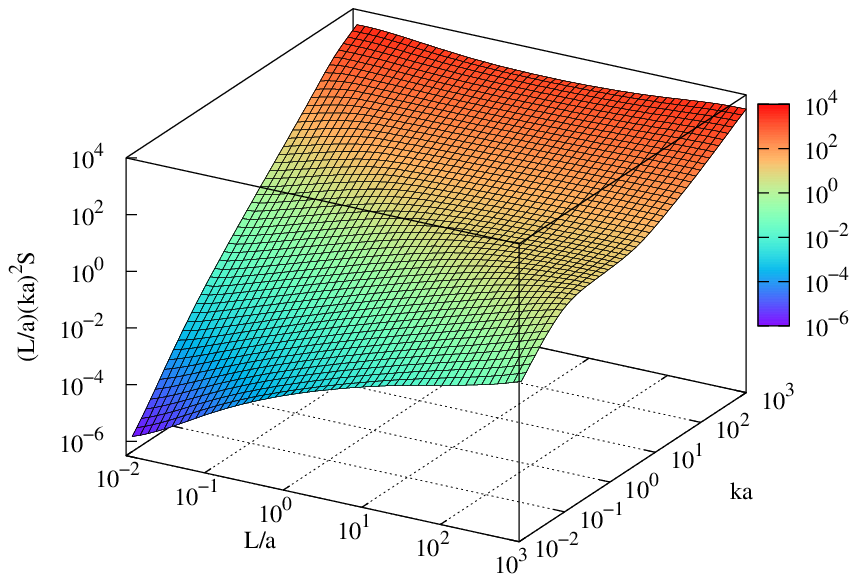}%
	\caption{\label{fig:fitting_result_3d} The continuous surface of $(L/a)(ka)^2S$ generated by the trained NN model .
	}
\end{figure}

What is more important is that our trained NN model can provide a continuous function of structure factor in entire $ L/a$--$ka $ space through the limited discrete training samples. As shown in Fig.~\ref {fig:fitting_result_3d}, a continuous $(L/a)(ka)^2S$ plane in logarithmic coordinates is obtained.  This result indicates that given \emph{any} $L/a$--$ka$ pair, the structure factor can be predicted directly. Also, we can easily calculate the values of \emph{multiple} structure factors \emph{simultaneously}, so the NN model greatly improves the calculation efficiency. 

\subsection{\label{sec:level5}The effect of network structure}

Hyperparameters, such as the number of neurons,  the number of layers,  optimizer, and learning rate, are also important in the training.  In this work, we focused on the effect of the number $H$ of hidden layers and the number $N$ of neurons in each hidden layer. 

To simplify the parameter adjustment process, we have made the number of nodes on each hidden layer the same. To study the effect of $ H $ on the fitting results, we fixed $ N $,  then we used four values of $ H$ ($1,2,4,8$)  to get four network structures. Finally, we have separately trained the NNs.

\begin{figure}
	\includegraphics{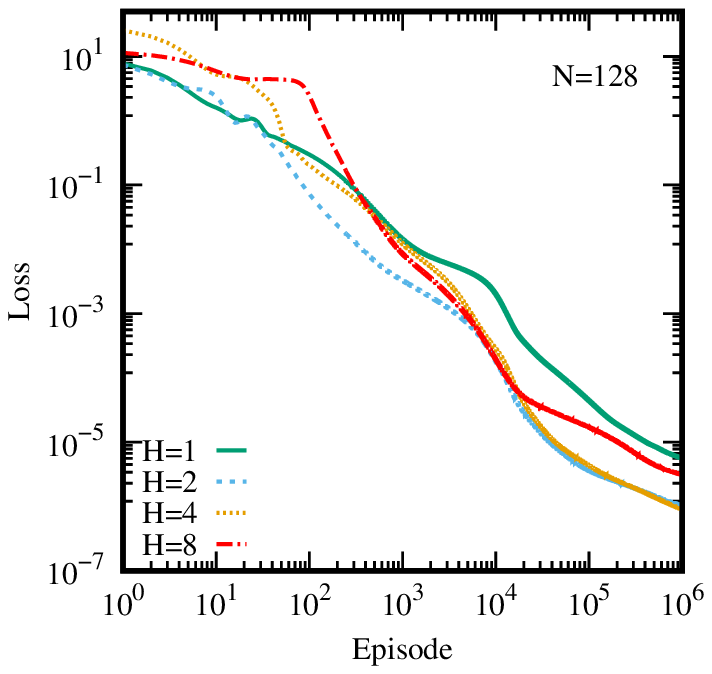}
	\caption{\label{fig:loss_change} Time dependence of the loss function for different number $H$ of hidden layers ($N=128$). }
\end{figure}

Fig.~\ref {fig:loss_change} shows the change of $ Loss $ in four independent pieces of training when the total number $N$ of hidden layer nodes is fixed at $ 128 $. At the end of training, episode $ \sim 900000$, the $ Loss $ for $ H = 2,4 $ is less than that for $ H = 1,8 $. This result indicates that when $N$ is fixed, too many or too few hidden layers will lower the performance of the trained NN model. Therefore, choosing the appropriate $ H $ can help make the model converge faster.

\begin{table}[b] 
	\caption{\label{tab:NH_Impact}
		The comparison of $Loss$ for different  $N$ and $H$ after $9\times10^5$ episodes of training.
	}
	\begin{ruledtabular}
		\begin{tabular}{ccccc}
			\textrm{\ }&
			\textrm{$N=32$}&
			\textrm{$N=64$}&
			\textrm{$N=128$}&
			\textrm{$N=256$}\\
			\colrule
			$H=1$   & $1.531\times10^{-5}$ &$8.059\times10^{-6}$ & $5.607\times10^{-6}$  & $7.129\times10^{-6}$\\
			$H=2$ &  $1.423\times10^{-6}$  &  $6.197\times10^{-7}$ & $1.024\times10^{-6}$  & $1.644\times10^{-6}$\\
			$H=4$ & $3.847\times10^{-6}$  & $1.680\times10^{-6}$ & $9.017\times10^{-7}$  & $5.411\times10^{-7}$\\
			$H=8$ & $2.597\times10^{-4}$ & $1.918\times10^{-5}$ & $3.109\times10^{-6}$  & $2.005\times10^{-6}$\\
			
		\end{tabular}
	\end{ruledtabular}
\end{table}

To further test the effect of $ N $, we trained the NN with $ N = 32,64,128 $ and $ 256 $, respectively. In Appendix.~\ref {appendix:hyperparameters},  the corresponding episode dependence of the  $ Loss $ for different $N$ are shown.  In Table.~\ref {tab:NH_Impact}, we listed the $ Loss $ values at the end of training when $ episode = 900000 $, with different $N$ and $H$. From the table, we noticed that for all $ N $, the $Loss$ for $ H = 2, 4 $ are smaller than  that for  $ H = 1, 8 $. Therefore, there must be an optimal H value between  $[H_{\text{min}}, H_{\text{max}}]$, which is $[1,8]$ in our case.  In addition, as $N$ increases, the loss value decreases more, which means the training processes can converge faster and can get more accurate prediction results.

\section{\label{Sec.2.P}Predict  the contour length and Kuhn length of polymer chains}

Small-Angle Neutron Scattering (SANS) is a widely used technique to study the structure of polymers. In SANS, the scattering intensity $I$ is measured as a function of the length of the scattering vector $q$. The structure factor of wormlike chains in NN formation developed in present work is an exact formation. Any previous approximation formations used to analyze the scattering experiments can be replaced by the NN formation directly.  In this part, we used some public scattering intensity data of polymers from the SANS experiment as examples to demonstrate the uses of  the  trained NN model, and then predicted the two important parameters of polymer chains, the contour length $L$ and Kuhn length $a$.

\subsection{Method} \label{relation_I_S}
The scattering intensity of a polymer chain can be fitted using the following equation\cite{Pedersen1996a,Hansen2013}
\begin{equation}\label{eq:I_p}
I_p(q) = cP(q)S(q)
\end{equation}
where $c$ is a scaling factor and $S(q)$ is structure factor obtained by the NN model, and 
\begin{equation}
P(q) = \left[ \frac{2J_{1}(Rq)}{Rq}\right]^{2}
\end{equation}
is the form factor, where $J_{1}(x)$ is the first order Bessel function and $R$ is the cross-section radius.  We approximated the finite cross section of the polymer chains into a cylinder with a radius of $R$ and a length of $a$.\cite{Pedersen1996a} The structure factor $S$ is determined by the parameter $L$ and $a$, and the form factor $P$ is determined by $R$.  Thus there are four fitting parameters: contour length $L$, Kuhn length $a$, radius $R$, and the scaling factor $c$.
 
We can use Eq.\ref{eq:I_p} to fit the SANS data by changing the parameters $L, a, R, c$. Define
\begin{equation}\label{eq:fit_app_eps}
\epsilon (a, L, R,c) = \frac{1}{N}\sum_{i=1}^N\left(I_p^{i}(a,L,R,c) - I^i \right)^2
\end{equation}
as the optimizing target,\label{fitting error} in which $I$ is the scattering intensity of a polymer chain obtained by the SANS and $I_p$ is the scattering intensity predicted by our NN model. The parameters $a, L, R, c$ need to be adjusted to make the predicted scattering intensity $I_p(q)$ and the measured one $I(q)$ as close as possible. When $\epsilon$ is small enough, the optimal parameters $L^{*}, a^{*}, R^{*}, c^{*}$ can be obtained. Therefore, we predicted the contour length ($L^{*}$) and Kuhn length ($a^{*}$) of the polymer chain.

There are multiple fitting parameters in the formula Eq.\ref{eq:fit_app_eps}, which will bring difficulties to the fitting of experimental data, especially in scattering data with fluctuations. The approximated structure factor of the chain in solution, Eq.\ref{eq:fit_app_eps}, is approximated by modified the ideal chain structure factor. To describe the effects of chain thickness and the solvent-monomer and monomer-monomer interaction etc,  some additional parameters have to be introduced in the formulation. There are many approximated formulations used in the scattering experiments in the literatures\cite{Pedersen1996a, Kholodenko1992, Mcculloch2013}. The prerequisite of these formulations is the structure factor of the ideal wormlike chain model. The major purpose of the present work is to develop an accurate structure factor formulation of the ideal wormlike chain model.

\subsection{Discussions}

Using the method in \ref{relation_I_S},  given the SANS intensity data of polymer chains, we can determine the contour lengths and Kuhn lengths. Two examples are given below.

\subsubsection{Polystyrene}
\begin{figure}
	\includegraphics{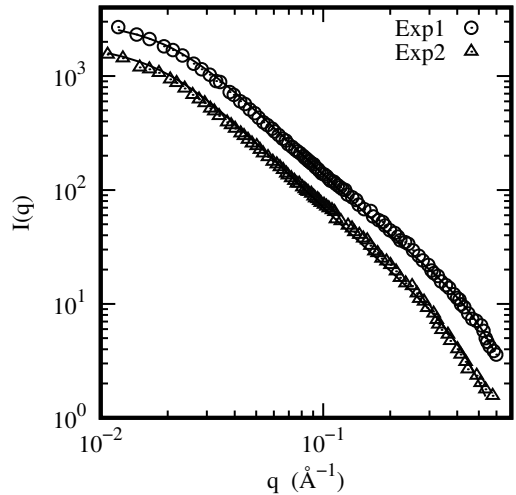} %
	\caption{\label{fig:predict_L_and_a} Scattering intensities comparison between the SANS data and the trained NN model prediction for PS with molecular wight $M_{w}$ of 50000 in CS$_2$. The Exp1 is for the phenylring deuterated PS, and the Exp2 is for fully deuterated PS.
	}
\end{figure}

The scattering intensity data of atactic polystyrene (PS) in carbon disulfide (CS2) with different selective deuteration of the polymer have been determined by Rawiso, Duplessix, and Picot \cite{Rawiso1987} using SANS. As shown in Fig.~\ref{fig:predict_L_and_a},  we have used two sets of scattering intensity data sets for the phenyl ring deuterated(Exp1: circle dots) and fully deuterated (Exp2: triangle dots) PS.  In Table~\ref{tab:ps_L_a}, the Kuhn lengths $a$ are determined to $ 22.38$ and $22.17$ \AA \   respectively , which are in good agreement with the previous determinations for SANS data($ 22\sim27$ \AA). And the contour lengths $L$ are $1300$ and $1574$\ \AA\ for Exp1 and Exp2 respectively. These results are also in good agreement with the values $1360$ and $1810$\ \AA \ in \cite{Pedersen1996a}. 

\begin{table}[b]
	\caption{\label{tab:ps_L_a}
		Fig.~\ref{fig:predict_L_and_a} 
	The predictions of $L$ and $a$ (unit: \AA)for PS.
	}
\begin{ruledtabular}
	\begin{tabular}{lcd}
		\textrm{\ }&
		\textrm{Upper}&
		\textrm{Lower}\\
		\colrule
		Reasonable Target Value  $L$ \footnote{These values was calculated from molecular weights and the structure taking into account the polydispersity.\cite{Pedersen1996a}} & $1360$  & $1810$ \\
		NN $L$ &  $1300.03$  &  $1573.76$\\
		Reasonable target value $a$ &  $22\sim27$  & 22\sim27 $$\\
		NN $a$ & $22.38 $ & $22.17 $\\
		$\epsilon$ \footnote{$\epsilon$ is fitting error defined in \ref{fitting error}} & $0.00068$ & $0.0013$ \\
	\end{tabular}
\end{ruledtabular}
\end{table}

\subsubsection{Poly(3-(2'-ethyl)hexylthiophene)}
Another set of  scattering intensity data of polymer chain P3EHT4 is from the experiments of Bryan McCulloch et. al.\cite {Mcculloch2013} Note that they proposed a very novel model of SANS intensity, the polydispersity-corrected wormlike chain model
\begin{equation} \label{eq:2013_I_model}
I(q)=K \int_{n=0}^{n=\infty} w_{i} g\left(u_{n i}\right) n_{i} \mathrm{d} n_{i}+I_{\mathrm{inc}}
\end{equation}
where the structure factor is denoted by $g$ in \cite{Mcculloch2013}

\begin{eqnarray}\label{eq:g}
g(u) \nonumber
&&=\frac{2}{u^{2}}\left(u-1+\mathrm{e}^{-u}\right)+ \\
&&\frac{2}{5 q^{2} L^{2}}\left[4 u-11 u \mathrm{e}^{-u}+7\left(1-\mathrm{e}^{-u}\right)\right], 
\end{eqnarray}
 $w_i$ is the weight fraction at a particular molecular weight, and 
 \begin{eqnarray}
 \nonumber
 u=q^{2} R_{\mathrm{g}}^{2}=q^{2}\left[\frac{L l_{\mathrm{p}}}{3}-l_{\mathrm{p}}+\frac{2 l_{\mathrm{p}}^{3}}{L}\left(1-\frac{l_{\mathrm{p}}}{L}+\frac{l_{\mathrm{p}}}{L} \mathrm{e}^{-L / l_{\mathrm{p}}}\right)\right],
 \end{eqnarray}
 where $l_p$ is the persistence length.

 In principle, we can directly use the structure factor $S$ obtained by our trained NN model to replace $g$ in Eq.\ref{eq:2013_I_model}, and then fit the contour length and Kuhn length of P3EHT4.  However, due to the lack of the original absolute molecular weight distribution of P3ETH4, we did not use Eq.~\ref{eq:2013_I_model} to  fit the SANS intensity data in this work. Alternatively, we have made a comparison chart of $ (L / a) (ka) ^ 2S $ and $ (L / a) (ka) ^ 2g $ with Kuhn length $ a = 10 $, to compare the structure factor $S$ from NN model and $g$ in Eq.~\ref{eq:g}.
 \begin{figure}
 	\includegraphics{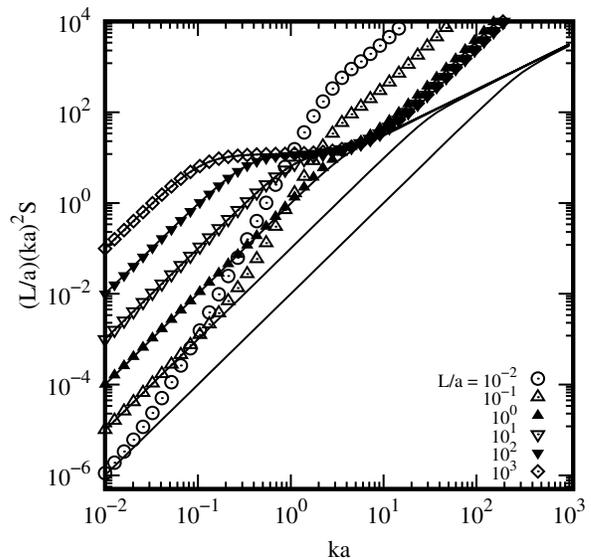}%
 	\caption{\label{fig:sg_compare}  The structure factor comparison between trained NN model (lines) and the wormlike chain model $g$ (circles, triangles, and etc.)\cite{Mcculloch2013}. 
 	}
 \end{figure}
And we found when  $ L / a> 1 $, $ ka <10 $, the $ S $ of the NN model and the $ g $ in Eq.~\ref{eq:g}  matched well as shown in Fig.~\ref {fig:sg_compare}.  Specifically,  because $g$ is deduced from the formula of the flexible chain model and uses an approximate form at large $k$.  Therefore, for flexible polymer chains($L/a=10^1, 10^2, 10^3$), $g$ gives a good  description of structure factor when $ka<10^0$.  But for large $ka$ ($ka>10^1$), $g$ describes the chain not very well.  In addition, for semi-rigid polymer chains ($L/a=10^0$),  $g$ is good when $k$ is small, but it is not good enough when $k>10^0$.  In fact, the semi-rigid chain is the key in many cases.  Besides, $g$ can not describe the structure factor of rigid chains($L/a<10^{-1}$), because the approximation for the rigid chains is not good enough.  The origin of $g$ here and the structure factor put by  Pedersen\cite{Pedersen1996a} and Kholodenko\cite{Kholodenko1992} are very similar. But their structure factor models can provide a better description in the rigid limit and the large $k$ limit.
 
As described in  section \ref{sec:level1},  our NN model can give precise predictions of  $S$ in the entire $L/a$--$ka$ space. Therefore, we expect the same fitting results as \cite{Mcculloch2013}, if $g$ in the Eq.~\ref{eq:2013_I_model}  is replaced by $S$. 
 Nevertheless, the Kuhn length $a$ of P3EHT4 determined by the intensity model we used in Eq.\ref{eq:I_p}  is $5.013$ \AA\  with $\epsilon$ minimized. 
 
\begin{figure}
 	\includegraphics{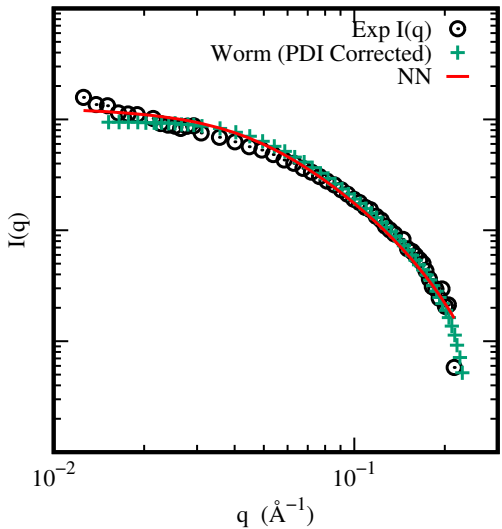}%
 	\caption{\label{fig:predict_2013} Scattering intensities comparison between the SANS data and the prediction of different model including Debye, Wormlike chain(PDI Corrected)\cite{Mcculloch2013} and the NN model for P3EHT-4.
 	}
 \end{figure}

 As shown in  Fig.~\ref{fig:predict_2013}, the intensity $I_p(q)$ calculated from our NN model  fits very well with  the SANS intensity $I(q)$.
 
\section{Summary}

We have developed an efficient model for the structure factor of a wormlike chain polymer by training a fully connected NN.  Our NN model is of the following characters: (a) High-precision, continuous numerical solutions in the entire $L/a$--$ka$ space can be obtained easily; (b) It is highly consistent with the calculations in previous numerical and analytical method\cite{Zhang2014}.  Besides, we also proposed one application of the model. Combining SANS intensity data we can determine the contour length and Kuhn length of polymer chains. Therefore, our NN model may provide a potential tool for exploring the properties of polymer chains for experimental researchers.

\appendix
\section{Structure factor obtained by interpolation}

Due to the monotonicity of the $(L/a)(ka)^2S$-surface, the structure factor may also be obtained by using suitable interpolation algorithms. In addition to using the NN, we also use two interpolation algorithms to accelerate the computation of the structure factor. In these two interpolation algorithms, we use the \emph{same} data points as described in Sec. \ref{subsec:IntroNN}.   

\begin{figure}[b]
	\centering 
	\subfigure[]
	{
		\includegraphics{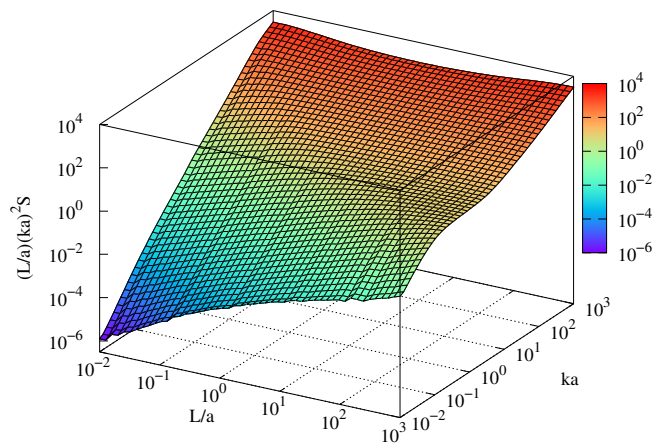}
		\label{fig:interpolation_nearest}
	}
	
	\subfigure[]
	{
		\includegraphics{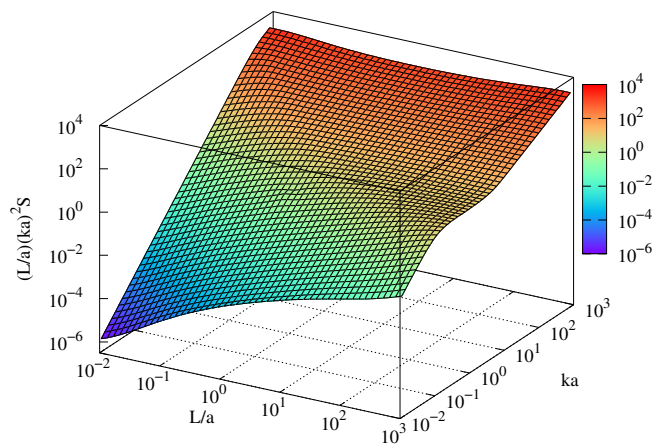}
		\label{fig:interpolation_cubic}
	}
	\caption{The surface of $(L/a)(ka)^2S$ obtained by interpolation. (a) Nearest-neighbor method; (b)  Cubic-spline method.} 
	\label{fig:interpolation}
\end{figure}

We found that interpolation algorithms approximately give the numerical solution of the structure factor in the entire $ka$, $L/a$ space. As shown in Fig.\ref{fig:interpolation}, (a) uses the nearest-neighbor method, (b) uses the cubic-spline method. The $(L/a)(ka)^2S$-surface obtained by the nearest-neighbor method in Fig. \ref{fig:interpolation_nearest} is less smooth than that by the cubic-spline method in Fig. \ref{fig:interpolation_cubic}. And the structure factor surface obtained by the cubic-spline method is closer to the solution of the MDE and the NN model. The interpolation works well at the small and large $k$ limits where the fractal dimension of wormlike chains can be well determined. For the medium range $k$ condition and semiflexible chain condition, $S$ varies rapidly. More data are required for the interpolation. 

\section{Loss Fuction For Different $N$}
\begin{figure*}
	\includegraphics{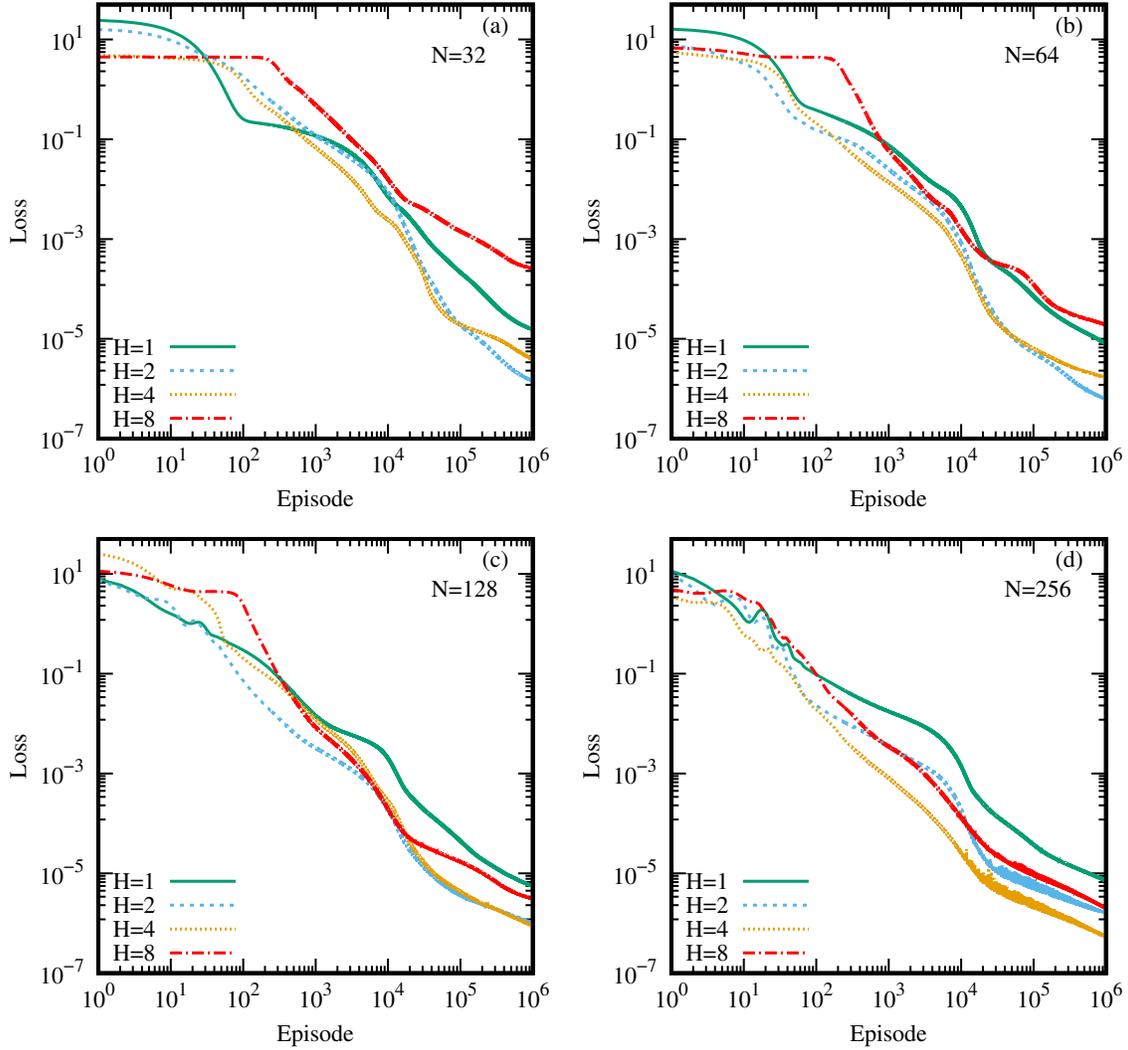}
	\caption{\label{fig:loss_change_big}Time dependence of the loss function for different $H$ and $N$ of the NN.  (a) $N=32$; (b) $N=64$; (c) $N=128$ and (d) $N=256$.}
\end{figure*}

Fig.\ref{fig:loss_change_big} shows how the loss function changes for different $N$.

\label{appendix:hyperparameters}

\section{Structure Factor}
In this appendix, we list some analytical expressions of the structure factor in different methods.
\subsection{Kholodenko}
In \cite{Kholodenko1993a},  the structure factor $S$ is obtained which correctly reproduces the rigid-rod and random-coil limits and is given analytically by
\begin{equation}
S(k)=\frac{2}{x}\left[I_{(1)}(x)-\frac{1}{x} I_{(2)}(x)\right]
\end{equation}
where $I_{(\mathrm{n})}(x)=\int_{0}^{x} f(z) z^{n-1} \mathrm{d} z$, $n=1,2$, $x=3L/a$,

\begin{equation}
f(z)=\left\{\begin{array}{ll}\frac{1}{E} \frac{\sinh (E z)}{\sinh z} & (k \leq 3 / 2 \mathrm{a}) \\ \frac{1}{\hat{E}} \frac{\sin (\hat{E} z)}{\sinh z} & (k>3 / 2 \mathrm{a})\end{array}\right.
\nonumber
\end{equation}

and 

\begin{equation}
E=\left[1-\left(\frac{2}{3} a k\right)^{2}\right]^{1 / 2},  \hat{E}=\left[\left(\frac{2}{3} a k\right)^{2}-1\right]^{1 / 2}.
\nonumber
\end{equation}

\subsection{Pederson and Schurtenberger}
In \cite{Pedersen1996a} , the structure factor of a semiflexible chain is given by
\begin{equation} \label{PS_main}
S=S_{\mathrm{SB}} P+S_{\mathrm{loc}}(1-P)
\end{equation}
where 

\begin{equation} \label{PS_loc}
S_{\mathrm{loc}}=\frac{c_{1}}{L a q^{2}}+\frac{\pi}{L q}
\end{equation}
is the approximate scattering function at high $q$ suggested by Burchard and Kajiwara\cite{Society1970},

\begin{equation} \label{PS_sb}
S_{\mathrm{SB}}=S_{\mathrm{Debye}}+\frac{c_{2} a}{L}\left[\frac{4}{15}+\frac{7}{15 x}-  \left(\frac{11}{15}+\frac{7}{15 x}\right) \exp (-x)\right]
\end{equation}
is the scattering function calculated for the Daniels approximation by Sharp and Bloomfield\cite{Sharp1968}, and 
\begin{equation}
P=\exp \left[-\left(\frac{q a}{q_{1}}\right)^{p_{1}}\right]
\nonumber
\end{equation}
where $q_1$ and $p_1$ are empirical constants. In Eq. \ref{PS_sb}, 
\begin{equation}
S_{\mathrm{Debye}}(x)=\frac{2}{x^{2}}[\exp (-x)+x-1]
\nonumber
\end{equation}
is the scattering function given by Debye function\cite{Debye1947a}, with $x \equiv R_{g}^{2} q^{2}$, and
\begin{equation} 
R_{g}^{2}=\frac{L a}{6}\left\{1-\frac{3 a}{2 L}+\frac{3 a^{2}}{2 L^{2}}-\frac{3 a^{3}}{4 L^{3}}\left[1-\exp \left(-\frac{2 L}{a}\right)\right]\right\}.
\nonumber
\end{equation}

The parameters depend on the $L/a$. 
For $L/a > 2$,  
$ c_{1} =1 ,  c_{2} =1 , p_{1} =5.33 , q_{1} =5.53  , R_{g}^{2} =La/6$. 
For $L / a \leq 2$, 
$c_{1} = 0.0625,  c_{2} = 0, p{1} = 3.95, q_{1} = 11.7a/L$.

\begin{acknowledgments}
	This research was supported by the Program of National Natural Science Foundation of China (NSFC) (Grant Nos. 21973070, 21774013, 21574011) and Beijing Natural Science Foundation (2182057). The authors also wish to express their appreciation for Jeff Z. Y. Chen for his valuable suggestions.  JH thanks Ying Jiang for his helpful guidance and discussion.
\end{acknowledgments}

\section*{Data availability statement}
The data that support the findings of this study are available from the corresponding author upon reasonable request.

\bibliography{nn_model_for_structure_factor}
\end{document}